\documentclass{blois}

\def\be{\begin{equation}}
\def\ee{\end{equation}}
\def\bea{\begin{eqnarray}}
\def\eea{\end{eqnarray}}

\newcommand{\Photo}{}

\begin{document}
\title{Probing T and CP Violation at DUNE and T2HK}

\renewcommand{\thefootnote}{\fnsymbol{footnote}}
\author{
 Sabya Sachi Chatterjee$^{a}$\,\footnote[1]{Speaker}\footnote[2]{sabya.chatterjee@kit.edu},
 Sudhanwa Patra$^{b,c}$,
 Thomas Schwetz$^a$,
 Kiran Sharma$^{a,b}$
}

\address{\vspace*{0.4cm}
$^a$ {\it {Institut f\"{u}r Astroteilchenphysik, Karlsruher Institut f\"{u}r Technologie (KIT), Hermann-von-Helmholtz-Platz 1, 76344 Eggenstein-Leopoldshafen, Germany}}\\
$^b$ {\it {Department of Physics, Indian Institute of Technology Bhilai, Durg-491002, India}}\\
$^c$ {\it
Institute of Physics, Sachivalaya Marg, Bhubaneswar- 751005, India}}

\maketitle

\abstracts{We study the sensitivity of the DUNE and T2HK long-baseline experiments to time reversal (T) violation in neutrino oscillations. Rather than the conventional approach of exchanging initial and final neutrino flavors, we search for T violation through the $L$-dependence of the $\nu_\mu \to \nu_e$ transition probability at fixed neutrino energy using neutrino data only. Within the standard three-flavour framework, we show that the DUNE and T2HK together can establish the presence of an $L$-odd component in the oscillation probability at up to $\sim 4\sigma$ significance, with the optimal sensitivity in the energy range $E_\nu \in [0.68, 0.92]$~GeV. The second oscillation maximum of DUNE plays a crucial role in this analysis. We further show that DUNE is more sensitive to T violation that is running in neutrino-only mode, whereas T2HK provides better sensitivity in the conventional neutrino versus anti-neutrino comparison, making the two experiments complementary to each other in search of the CP phase $\delta_{\rm CP}$.}

\section{Introduction}
%
The discovery of a non-trivial complex phase in the PMNS neutrino mixing matrix~\cite{Pontecorvo:1967fh} is one of the central goals of next-generation long-baseline experiments.  Within the standard three-flavour framework, CP and T symmetry violations in neutrino oscillations are equivalent and both governed by the single phase $\delta_{\rm CP}$.  The traditional approach for the T violation search requires exchange of initial and final neutrino flavours, which needs the advent of a neutrino factory and is experimentally very challenging. In this work we pursue a different route \cite{Schwetz:2021cuj,Chatterjee:2024jzt,Chatterjee:2025ssc}: we search for T violation through the \emph{$L$-dependence} of the $\nu_\mu \to\nu_e$ oscillation probability at fixed neutrino energy.
Under T conjugation the oscillation probability transforms as $P_{\alpha\to\beta}(L)\to P_{\alpha\to\beta}(-L)$.  Hence T violation manifests itself as an \emph{$L$-odd} component of the appearance probability, and we argue that a measurement at fixed neutrino energy and two different baselines with neutrino data only can establish its presence.  The DUNE ($L=1300$~km) and the T2HK ($L=295$~km) experiments are ideally suited for this, while both produce intense $\nu_\mu$ beams covering an overlapping energy region. For CP violation search we follow the conventional method of comparing the neutrino and antineutrino oscillation probabilities.
%
%
\section{Formalism}
%
We focus on the appearance probability
$P_{\nu_\mu\to\nu_e} = \left|\sum_{i=1}^3 c_i\,e^{-i\lambda_i L}\right|^2$, where $\lambda_i$ are the eigenvalues of the effective Hamiltonian in matter,
$c_i \equiv U^{m*}_{\mu i}\,U^m_{ei}$, and $U^m$ is the effective mixing matrix in matter.  The probability splits naturally into T-even and T-odd parts under $L \to -L$:
\vspace{-0.3cm}
\bea
  P_{\rm even} &=& 4|c_2|^2\sin^2\!\phi_{21}
    +4|c_3|^2\sin^2\!\phi_{31}
    +8\,\mathrm{Re}[c_2^*c_3]\sin\phi_{21}\sin\phi_{31}\cos\phi_{32}\,,\nonumber \\
  P_{\rm odd}  &=& 8\,\mathrm{Im}[c_2^*c_3]\sin\phi_{21}\sin\phi_{31}\sin\phi_{32}\,,
  \label{Eq:T_even_odd}
\eea
with $\phi_{ij}=\Delta m^2_{ij,\rm eff}(E_\nu)\,L/(4E_\nu)$. The anti-neutrino probability requires both complex conjugation of the $c_i$ and a sign change of the matter potential. T violation is established if the data cannot be described by $P_{\rm even}$ alone. CP violation requires the comparison of neutrino and antineutrino probabilities which, however, is contaminated by the ``environmental'' asymmetry induced by the sign-change of the matter potential.
%
%
%
\section{Experimental Setup}

{\bf DUNE}: We adopt the TDR configuration~\cite{DUNE:2020jqi} with a 40~kton LArTPC far detector at 1300~km, a 1.2~MW beam power delivering $1.1\times10^{21}$~POT/yr. Our default exposure is 336~kt\,MW\,yr, accumulated entirely in neutrino mode over 7 yrs running period. For {\bf T2HK}, we follow~\cite{Hyper-Kamiokande:2018ofw}, with a 187~kton water Cherenkov detector at 295~km, driven by the 30~GeV J-PARC proton beam delivering $1.1\times10^{21}$~POT/yr. The default exposure is 608~kt\,MW\,yr in neutrino mode ($\approx 2.5$~yr). Details about the energy resolutions, systematics uncertainties and statistical method used in the analyses can be found in~\cite{Chatterjee:2025ssc}. Throughout the analyses we have assumed normal ordering, line-averaged matter density $\bar{\rho} = 2.84\, g/cc$ for both the experiments and the following benchmark choices of the standard oscillation parameters~\cite{Esteban:2024eli}:
$\sin^2\theta_{12}=0.307$, $\sin^2\theta_{13}=0.022$, $\sin^2\theta_{23} = 0.561$, $\Delta m^2_{21} = 7.49$~[eV$^2$] and $\Delta m^2_{31} = 2.53$ [eV$^2$].
\section{Sensitivity to T Violation}
%
\begin{figure}[h]
 \centering
 \includegraphics[width=0.32\textwidth]{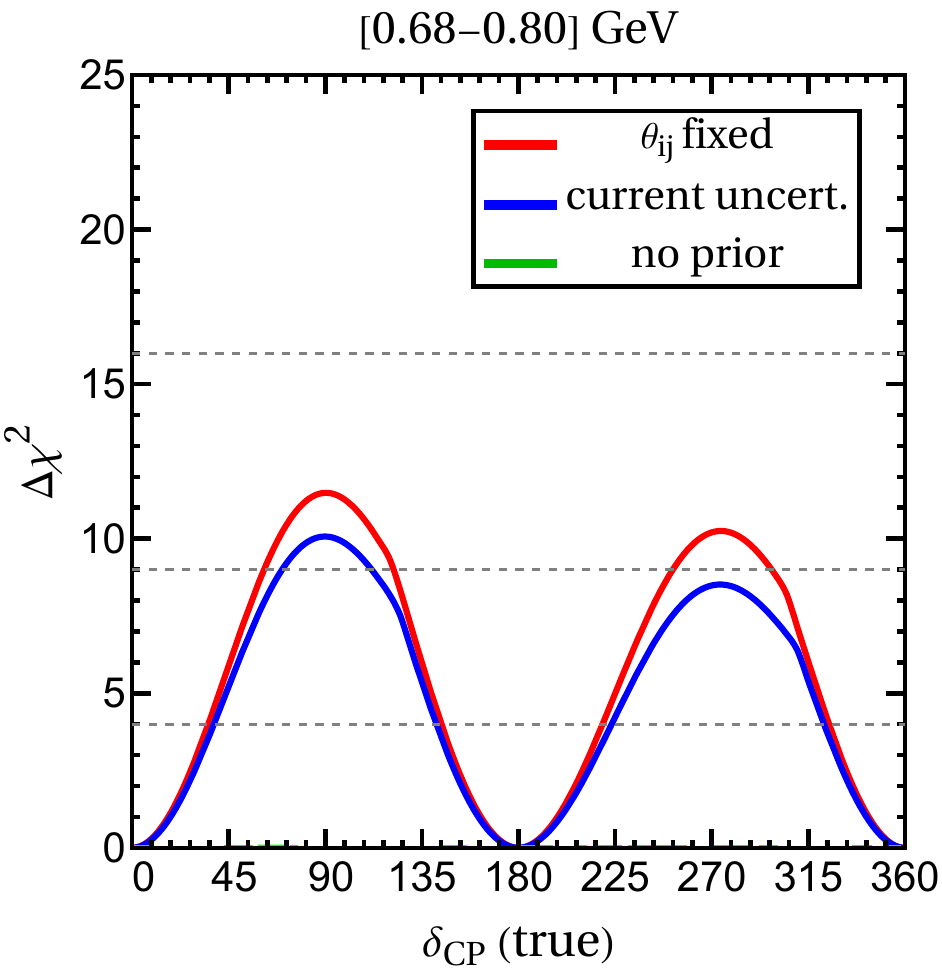}
 \includegraphics[width=0.32\textwidth]{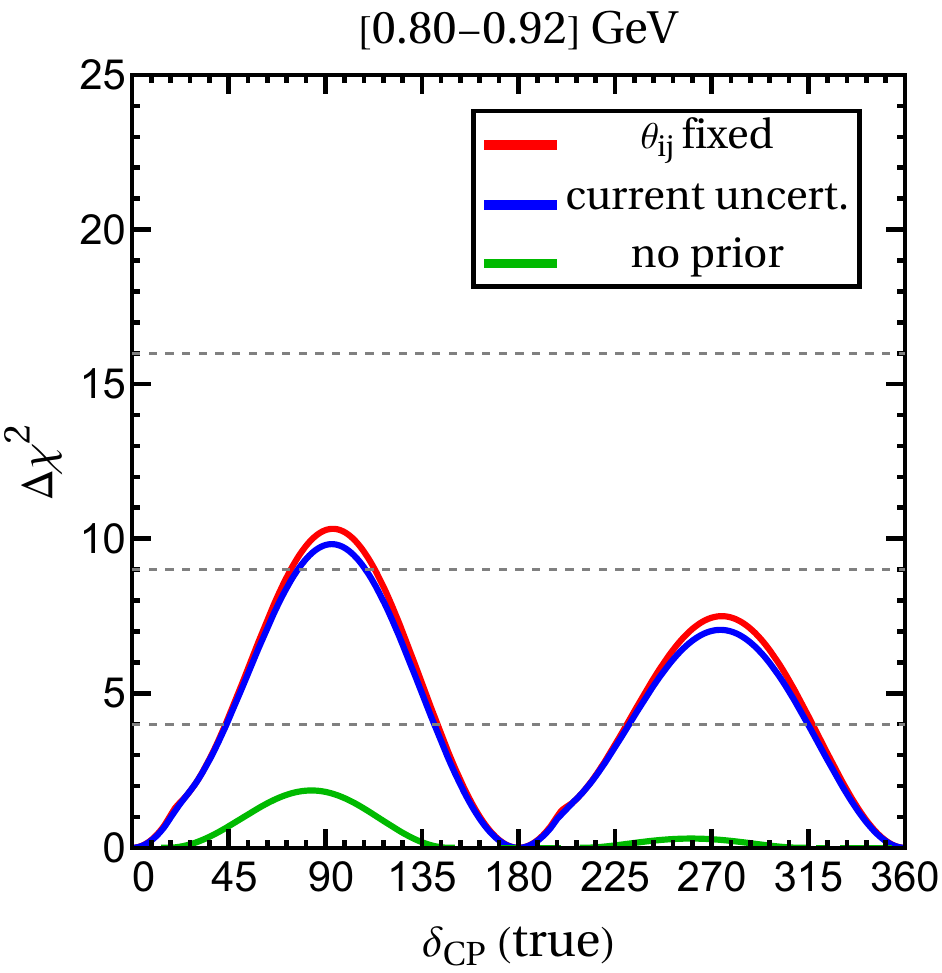}
 \includegraphics[width=0.32\textwidth]{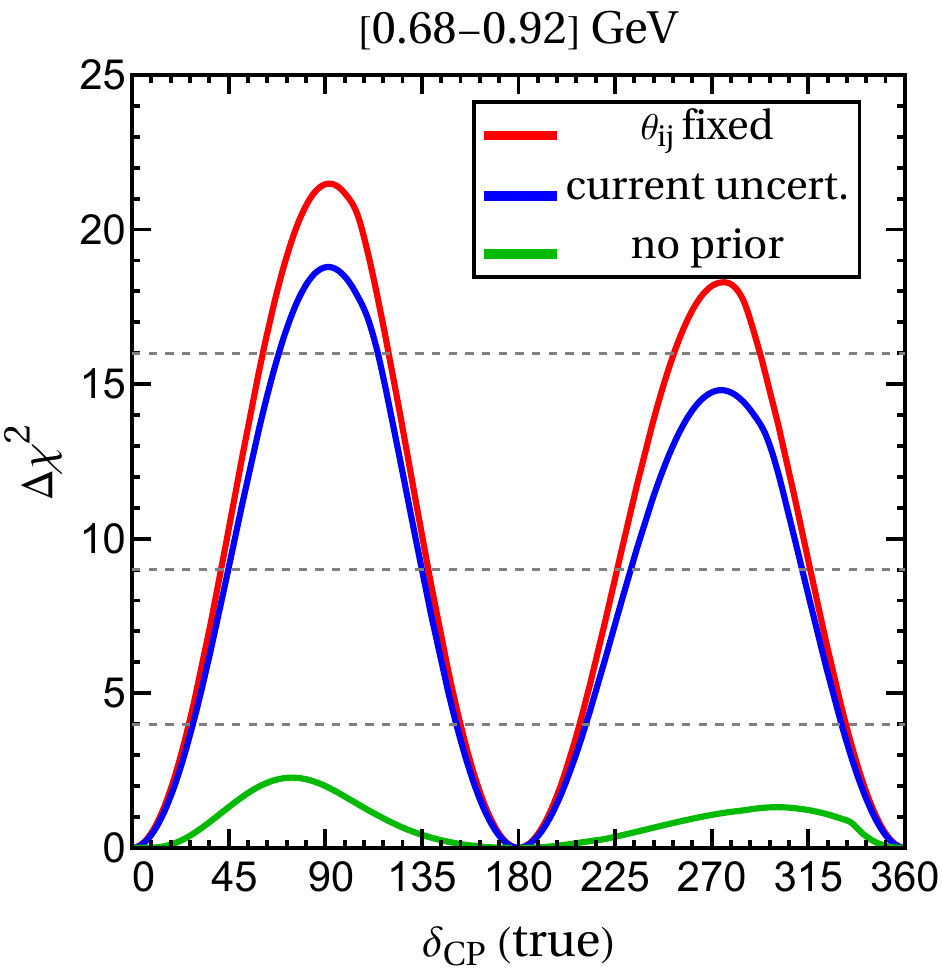}
 \caption[]{Sensitivity to T violation for DUNE+T2HK as a function of $\delta_{\rm CP}$ (true) for two energy bins in the range $[0.68,\,0.92]~ \rm GeV$ shown separately (left and middle panel) and combined (right panel). Different colors of the curves correspond to different prior assumptions on the mixing angles $\theta_{ij}$. This figure is taken from~\cite{Chatterjee:2025ssc}.}
 \label{Fig:Tsens_1}
\end{figure}
Fig.~\ref{Fig:Tsens_1} shows the sensitivity of T violation as a function of true $\delta_{CP}$ where we have assumed the neutrino mode only and default configurations for both DUNE and T2HK. We find that the energy range $E_\nu\in[0.68,0.92]$~GeV provides the best sensitivity, split into two bins of width 0.12~GeV each. The sensitivities are shown for the individual bins as well as in their combined mode assuming different choices of parameters as indicated by different colors in each plot. The combined analysis of the two bins in the right panel shows excellent sensitivity around $4\sigma$ for maximal T violating phases provided the mixing angles are well known. We have found that the good sensitivity to T violation in the energy bin $\left[0.68-0.80 \right]$~GeV emerges mainly from the synergy of the two experiments where T2HK has its spectral peak and DUNE covers its second oscillation maximum. However for the case of $\left[0.80-0.92 \right]$~GeV it is mainly dominated by DUNE. For more detailed explanation using the biprobability plots and for further details of more bin-wise sensitivities of individual experiments see~\cite{Chatterjee:2025ssc}.
\begin{figure}[h]
 \centering
 \includegraphics[height=7cm,width=7cm]{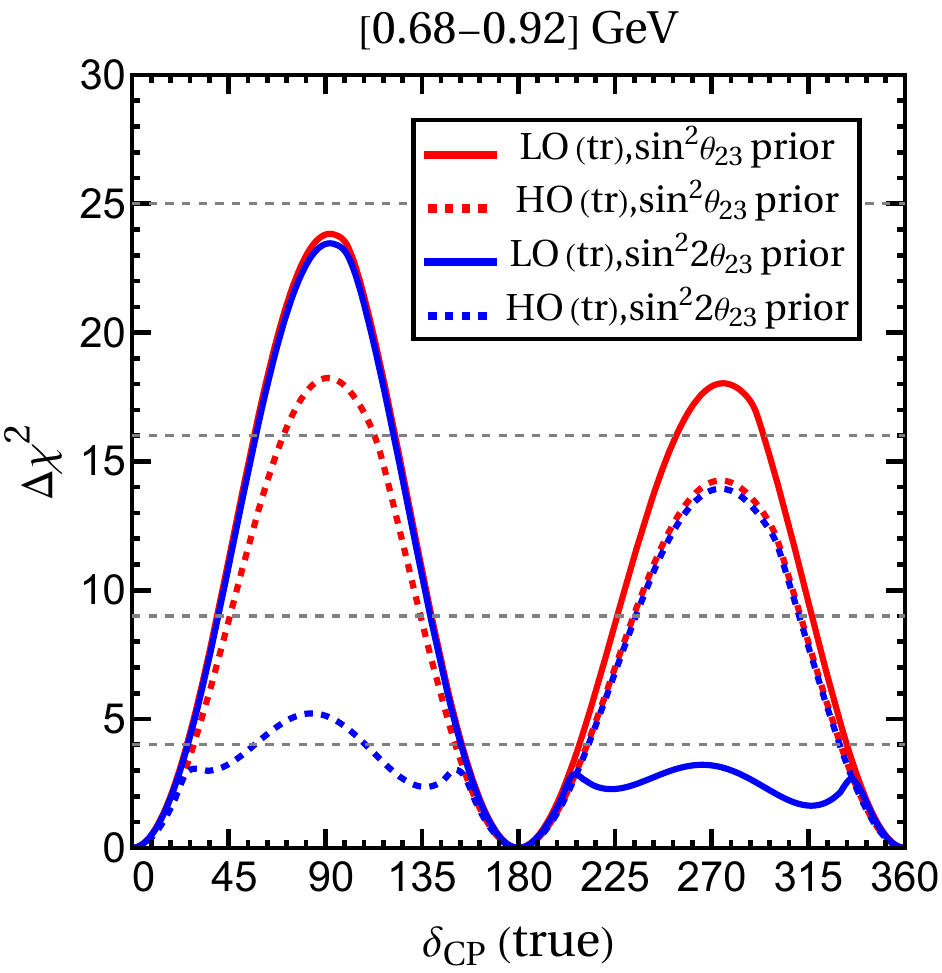}
 \includegraphics[height=7cm,width=7cm]{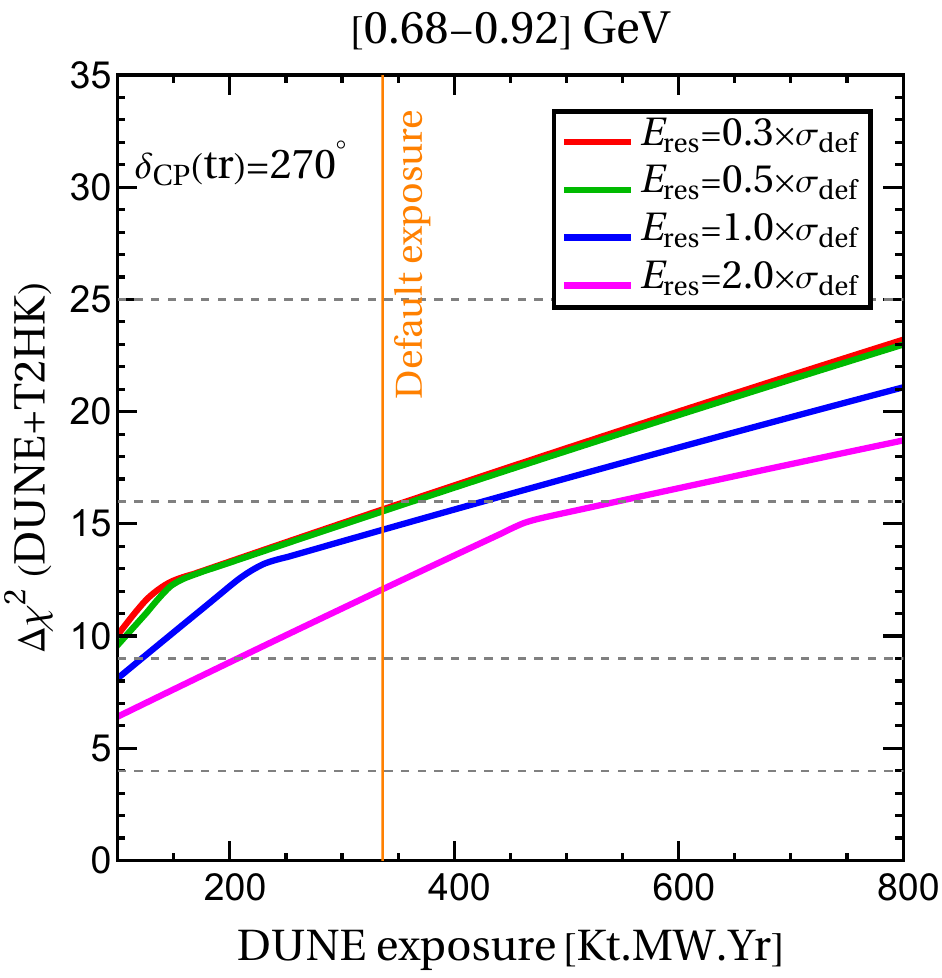}
 \caption[]{Left panel: impact of the $\theta_{23}$ octant on the T-violation sensitivity. Solid (dashed) curves assume $\sin^2\theta_{23} = 0.435\,(0.561)$, corresponding to the first (second) octant. Red (blue) curves use a prior on $\sin^2\theta_{23}$ ($\sin^2 2\theta_{23}$), with uncertainty $\sigma = 0.015\,(0.01)$. Right panel: sensitivity as a function of DUNE exposure for $\delta_{\rm CP}({\rm true}) = 270^\circ$, with T2HK fixed at 608~kt\,MW\,yr. This figure is adopted from~\cite{Chatterjee:2025ssc}.}
 \label{Fig:Tsens_2}
\end{figure}
Fig.~\ref{Fig:Tsens_1} highlights the importance of the prior knowledge of the mixing parameters in determing the T violation sensitivities. In the left panel of Fig.~\ref{Fig:Tsens_2} we discuss the impact of $\theta_{23}$ octant degeneracy on T violation search. By default we assume the benchmark value of $\sin^2\theta_{23}=0.561$ in our analysis and impose a prior on $\sin^2\theta_{23}$, which resolves the octant degeneracy. To assess how important the choice of prior is, we also consider a prior on $\sin^22\theta_{23}$, which is symmetric under octant exchange and therefore leaves the degeneracy unresolved. The left panel shows the impact of these choices in detail. The red solid (dashed) curve shows the result assuming prior on $\sin^2\theta_{23}$ for true choice of lower (higher) octant. Similarly the blue curves are shown assuming prior on $\sin^22\theta_{23}$. The comparison shows that knowing the octant matters considerably, especially for the combinations ($\theta_{23}>45^\circ$, $\delta_{\rm CP}\simeq90^\circ$) and ($\theta_{23}<45^\circ$, $\delta_{\rm CP}\simeq270^\circ$). Since both DUNE and T2HK have good sensitivity to determine the octant on their own, we take the prior on $\sin^2\theta_{23}$ as default, assuming the octant will be known by the time these data are analysed.

The two most optimized bins considered in the analysis lie in the low energy regime of DUNE spectrum where low event numbers are expected. Therefore, we study the sensitivity as a function of DUNE exposure in the unit of kt MW yr and our results are shown in the right panel of Fig.~\ref{Fig:Tsens_2}. We have fixed the T2HK to its default energy resolution and its exposure to 608 kt MW yr due to the negligible impact observed when other choices are made. The different color in each curve represents the different choices of the energy resolution of DUNE far detector. It is clear from the figure that for our default choice of the energy resolution (blue curve) the sensitivity reaches about $4\sigma$ at around 400 kt MW yr of exposure in the neutrino running mode for $\delta_{\rm CP}(\rm true) = 270^\circ$. For more details see~\cite{Chatterjee:2025ssc}.

\begin{figure}[h]
 \centering
 \includegraphics[height=7cm,width=7cm]{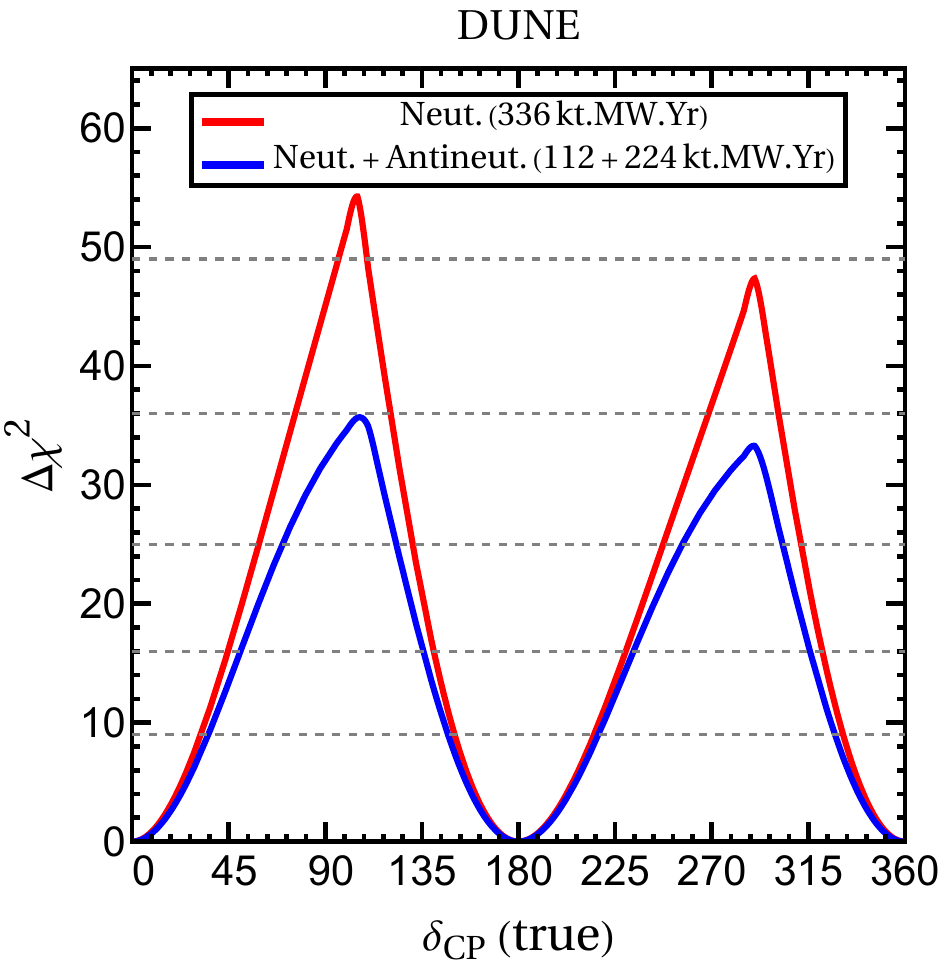}
 \includegraphics[height=7cm,width=7cm]{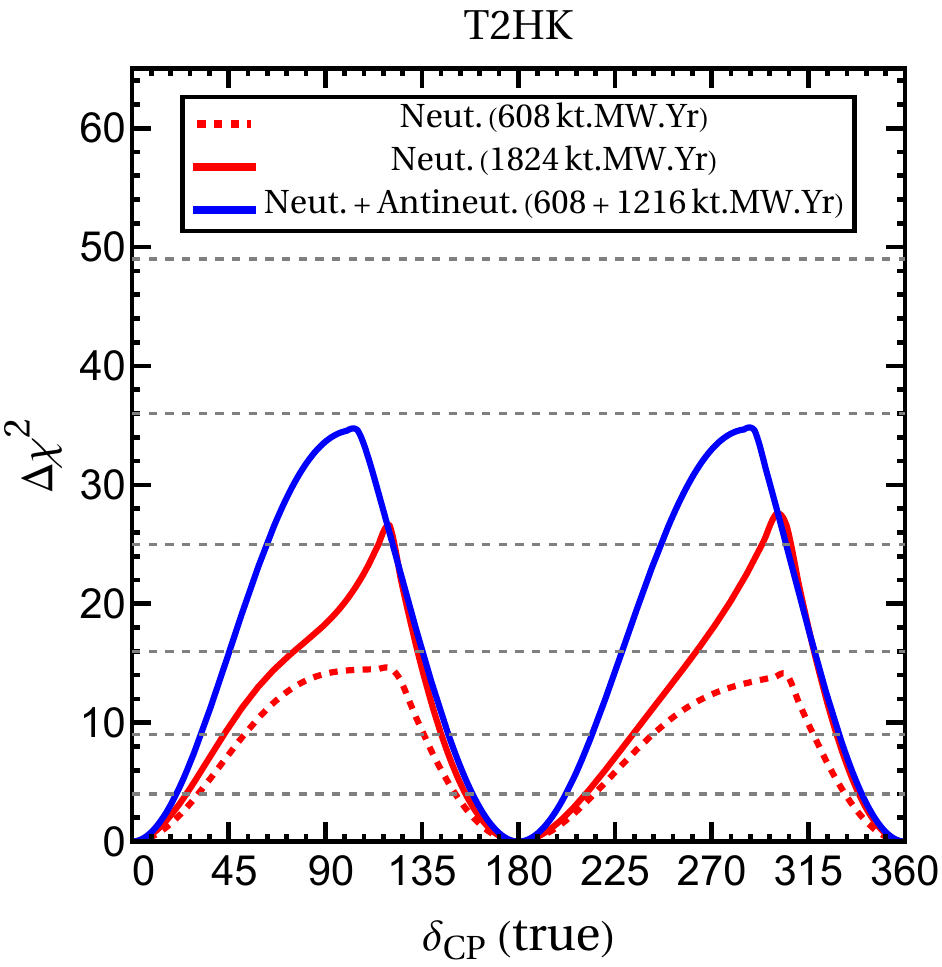}
 \caption[]{Sensitivity to $\delta_{\rm CP}\neq 0, \pi$ from a full spectral analysis for DUNE (336 kt MW yr) in the left and for T2HK (1824 kt MW yr) in the right. Red curves use the full exposure in neutrino mode only, while blue curves split it between neutrino and anti-neutrino running in the ratio 1:2. For T2HK, the neutrino-only sensitivity at 608 kt MW yr is also shown for comparison. This figure is taken from~\cite{Chatterjee:2025ssc}. }
 \label{Fig:T_vs_CP_sens}
\end{figure}
\section{T Violation versus CP Violation: Complementarity}

We contrast the T-violation approach (neutrino mode only) with the conventional CP-violation search (neutrino + anti-neutrino spectra). In both cases we adopt the method of fitting full energy spectra. For each experiment we compare two scenarios with the same total exposure: (i)~full exposure in neutrino mode and (ii)~exposure split 1:2 between neutrino and anti-neutrino beams.
The key result is that the two experiments are \emph{complementary}.  DUNE achieves better sensitivity in neutrino-only mode which can be attributed to the longer baseline and broad spectrum covering both first and second oscillation maxima. This means that the $L$-odd (T-violating) component carries the dominant information on $\delta_{\rm CP}$.  T2HK, on the other hand, achieves better sensitivity in the neutrino\,+\, anti-neutrino mode, because of the shorter baseline the first oscillation maximum dominates, which makes the CP-asymmetric observable more powerful.

\section{Conclusions}

We have shown that the DUNE and T2HK long-baseline experiments together can establish T violation at up to $\sim4\sigma$ significance by searching for the $L$-odd component in the $\nu_{\mu} \to \nu_e$ transition probability, using only neutrino beam data and the optimal energy window for that is $E_\nu\in[0.68,0.92]$~GeV, where the second oscillation maximum of DUNE plays a crucial role.  Prior knowledge of the mixing angles, especially the $\theta_{23}$ octant, is essential to reach $>3\sigma$ sensitivity.  Regarding the complementarity of the two experiments, DUNE is more sensitive to T violation that is running in neutrino mode only, while T2HK is more sensitive to CP violation from the neutrino versus anti-neutrino comparison. Finally we have shown that, for the default energy resolution of DUNE detector, a sensitivity of about $4\sigma$ can be reached at an exposure of around 400~kt MW yr in neutrino mode for $\delta_{\rm CP}(\rm true) = 270^\circ$.  A better resolution allows to reach the same significance even at smaller exposures. These results highlight the importance of optimising the low-energy performance of the DUNE detector for the T-violation search.
This analysis offers a new, physically transparent interpretation of future long-baseline data rather than merely fitting for $\delta_{\rm CP}$ within a model, one can directly observe the $L$-odd component of the oscillation probability that is the hallmark of fundamental T violation in neutrino mixing.

\bigskip

\textbf{Acknowledgments.}
The work of S.S.C.\ is funded by the Deutsche Forschungsgemeinschaft (DFG, German Research Foundation), project number 510963981.

\end{document}